\documentclass[aip,jcp,reprint,groupedaddress,twocolumn]{revtex4-1}
\usepackage{graphicx}
\usepackage{epsfig}
\usepackage{amsmath}
\usepackage{hhline}

\newcommand{\braket}[1]{\ensuremath{\langle{#1}\rangle}}

\newcommand{\beq}{\begin{equation}}
\newcommand{\eeq}{  \end{equation}}
\newcommand{\ba}{\begin{eqnarray}}
\newcommand{\ea}{  \end{eqnarray}}

\begin{document}

\title{Out-of-equilibrium catalysis of chemical reactions by electronic tunnel currents}
\author{Alan A.  Dzhioev}
\affiliation{Bogoliubov Laboratory of Theoretical
Physics, Joint Institute for Nuclear Research, RU-141980 Dubna, Russia}
\affiliation{Department of Physics, Universit\'e Libre de Bruxelles, Campus Plaine, CP 231, Blvd du Triomphe, B-1050 Brussels, Belgium}

\author{Daniel S. Kosov}
\affiliation{School of Engineering and Physical Sciences, James Cook University, Townsville, QLD, 4811, Australia}
\affiliation{Department of Physics, Universit\'e Libre de Bruxelles, Campus Plaine, CP 231, Blvd du Triomphe, B-1050 Brussels, Belgium}

\author{Felix von Oppen}
\affiliation{Dahlem Center for Complex Quantum Systems and Fachbereich Physik, Freie Universit\"at Berlin,
Arnimallee 14, 14195 Berlin, Germany}


\begin{abstract}
We present an  escape rate theory for current-induced chemical reactions.
We use Keldysh nonequilibrium Green's functions to derive a Langevin equation for the reaction coordinate. Due to the out of equilibrium electronic degrees of freedom, the friction,  noise,  and effective temperature in the Langevin equation depend locally on the reaction coordinate.
As an example, we consider the dissociation of  diatomic molecules induced by the  electronic current from a scanning tunnelling microscope tip.
In the resonant tunnelling regime, the molecular dissociation involves two processes which are intricately interconnected: a modification
of the potential energy barrier and heating of the molecule. The decrease of the molecular barrier (i.e.  the current induced catalytic reduction of the barrier) accompanied by the appearance of the effective, reaction-coordinate-dependent temperature is an alternative mechanism for current-induced chemical reactions, which is distinctly different from the usual paradigm of  pumping vibrational degrees of freedom.

\end{abstract}

\maketitle

\section{Introduction}

The recent advances in nano fabrication make it possible to use the scanning tunnelling microscope (STM)  as  a "nanoscale chemical reactor".
The tunnelling current from the STM tip can selectively break or form  chemical bonds
\cite{PhysRevLett.78.4410,PhysRevLett.84.1527,ho99}
and initiate  chemical reactions of the reactants.\cite{PhysRevLett.85.2777,Repp26052006}

The  rates of chemical reactions can be routinely computed for  molecular systems  in thermodynamic equilibrium.
Usually one relies on the Born-Oppenheimer approximation\cite{doi:10.1021/ar00028a007,doi:10.1021/j100238a003,uwe11} and may include various nonadiabatic effects, when the electronic energy levels are not well separated or coupled to the continuum of the environment states.\cite{tully:1061,kohanoff94,doltsinis02,tully95}
Let us now assume that the electronic system is driven out of equilibrium.
As an example for an out of equilibrium system we consider a molecular junction (molecule attached to two macroscopic metal electrodes held at different chemical potentials) or  a molecule absorbed on a metal surface under a scanning tunnelling  microscope tip.  As electric current is flowing through the molecule,  considerable amounts of energy are dissipated and partly passed from electronic to nuclear degrees of freedom --- in  linear response
the total dissipated power  is proportional to $IV$. Thus, since the current is in the range of nanoamperes and the voltage is a few volts,  a significant energy ($\sim 0.2$ Hartree per nanosecond) is dissipated in total and the energy which is dissipated in the molecule  per nanosecond can be comparable to typical barriers for chemical reactions.

Generally speaking, the electrons produce not only the standard adiabatic force but also give  random fluctuations and viscosity to the nuclear dynamics.\cite{tully95} Close to equilibrium, the latter are related by the fluctuation-dissipation theorem,\cite{dzhioev11} but  far from equilibrium  the noise is no longer balanced by the viscosity.
The absence of the fluctuation-dissipation relation as a  direct consequence of nonequilibrium opens
new possibilities in chemistry, such as  unusual ways to catalyse chemical reactions.

In this paper, we develop a theory for  current-induced chemical reactions. The   main physical assumptions of our theory are  that the electronic dynamics is much faster than the  nuclear motion and that the reaction coordinate is a classical variable. This results in Markovian dynamics of the reaction coordinate described by a Fokker-Planck equation. This Fokker-Planck equation has  interesting new features. Namely, the nonequilibrium potential energy surface depends on the electronic current flow through the molecule and the effective temperature produced by the nonequilibrium electrons on nuclear degrees of freedom is no longer constant but becomes a function of the reaction coordinate.
In other words, the out of equilibrium electrons not only change the profile of the potential energy surface for the reaction coordinate  but locally heat  it  with different rates.
For the paradigmatic  example of the dissociation of diatomic molecules, we show that  current-induced chemical reactions  involve two interconnected processes: a modification of the potential energy barrier and heating of the molecule. This  new mechanism for current-induced chemical reactions,  complements the familiar paradigm of  pumping vibrational degrees of freedom.\cite{vonoppen06,PhysRevB.83.115414}
The principal difference between this paper and previous work \cite{dzhioev11} lies in lifting the strong assumption that the temperature of the nuclear degrees of freedom is exactly the same as that of the equilibrium electrons in the metal surface.

The remainder of the paper is organized as follows. Our main results are
summarized and illustrated by Figs. 7 and we discuss our results in
relation to experiment at the end of Sec.~3. In Sec.~2, we describe the Langevin equation for the reaction coordinate.
 Section~3 presents the calculation of the reaction rates via the solution of a Fokker-Planck equation.
Conclusions are given in Sec.~4. Some technical aspects are relegated to  appendices.  We use atomic units throughout the equations in the paper.

\section{Langevin equation for the reaction coordinate}

\begin{figure}[t!]
\begin{center}
\includegraphics[width=0.8\columnwidth]{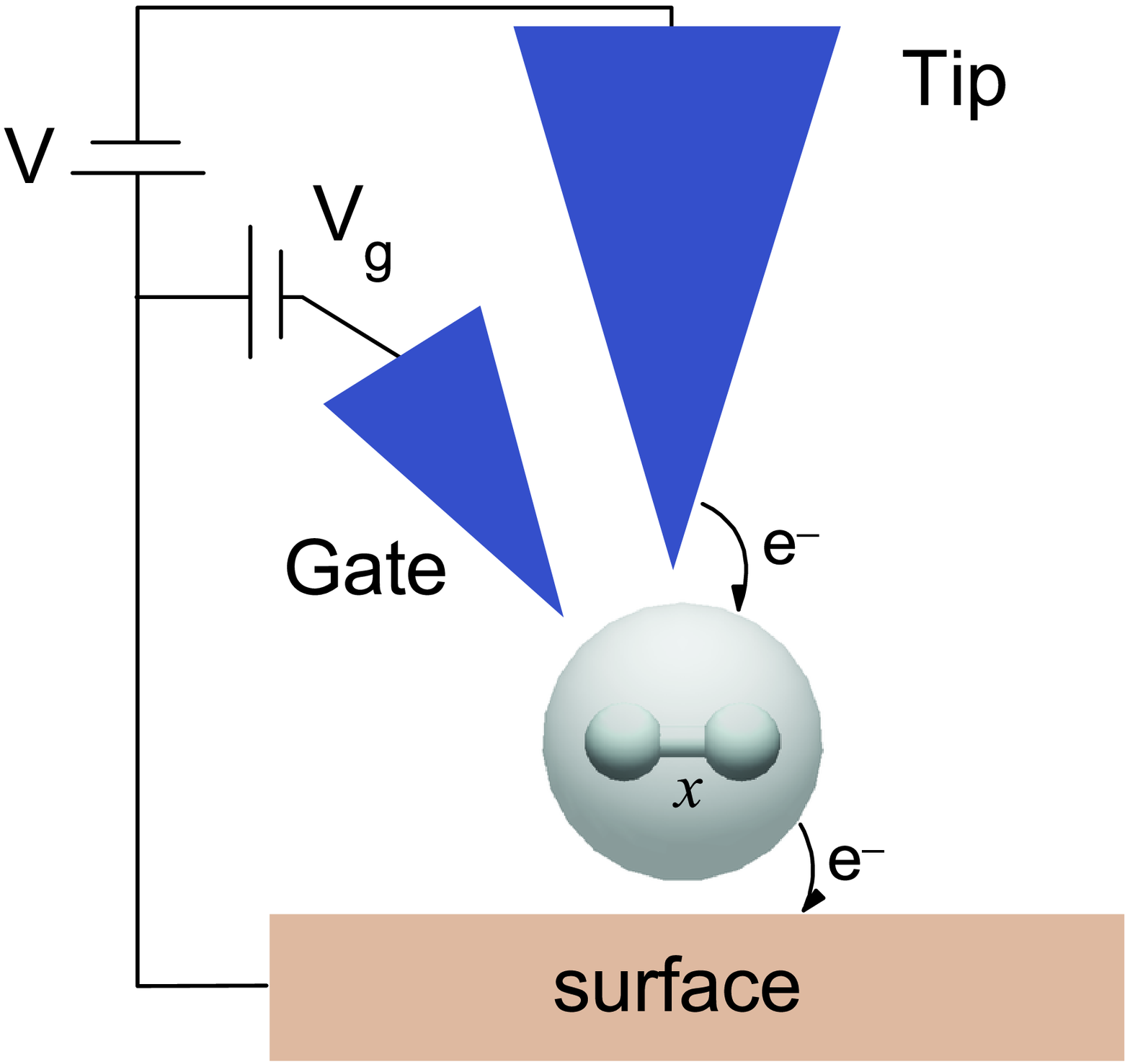}
\includegraphics[width=0.8\columnwidth]{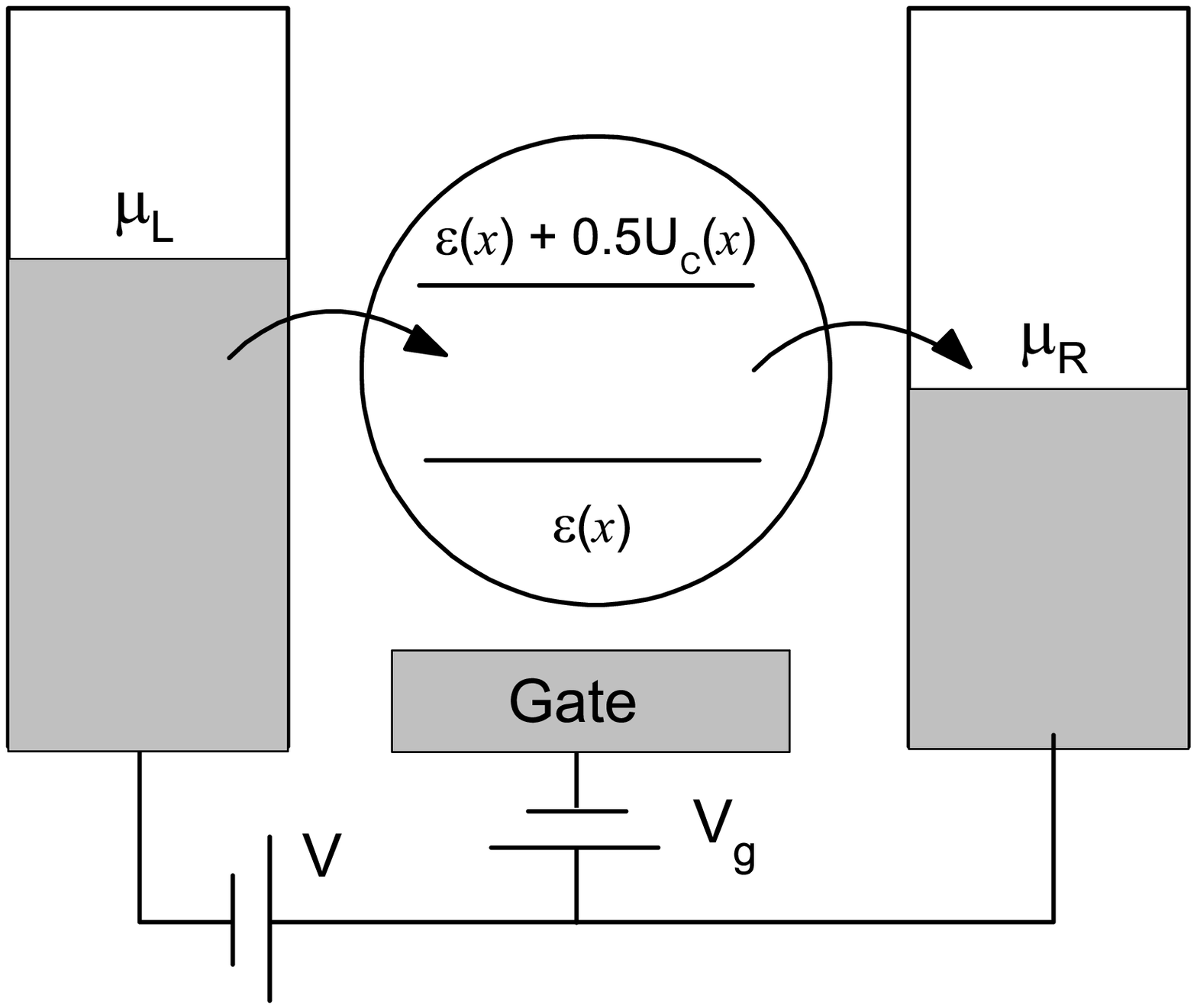}
\end{center}
\caption{Upper part: Possible experimental setup. The STM tip is positioned above a  molecule (represented as a sphere).
An electron coming from the tip onto the molecule then tunnel into the conducting surface. A gate electrode is used to tune
the molecular energy level(s) and to control the current flow between the STM tip and the conducting surface. Lower part: Schematic illustration of the model.
By varying the gate voltage the molecular energy level can be shifted near the chemical potential of the electrodes. The molecular orbital energy $ \varepsilon(x)$ and electron-electron interaction energy $U_\mathrm{C}(x)$ depend on the bond length~$x$.}
\label{H2}
\end{figure}

As a model system, we consider a diatomic molecule attached to two metal electrodes,
say, a metal surface on one side and an STM
tip on the other. A sketch of a possible experimental
setup is shown in Fig.~\ref{H2} .
The molecule is modeled by one electronic spin-degenerate molecular orbital
with single particle energy $\varepsilon(x)+V_g$, which depends on the bond length    $x$ and the gate voltage $V_\mathrm{g}$.
The bond length $x$ can be considered as the reaction coordinate.  The Coulomb interaction between electrons
is accounted for by a charging energy $U_\mathrm{C}(x)$ which is a function of the bond length $x$.
The reaction coordinate $x$  is considered to be a classical variable with corresponding momentum $p$
and  a reduced mass $m$ (taken as that of  the H$_2$ molecule, $m=918$ a.u.).
The nuclear Coulomb repulsion energy  is  $V_{N}(x)=1/x$.
Then the  molecular Hamiltonian is
\begin{equation}
H_M = (\varepsilon(x)_\mathrm{}+V_g) \sum_\sigma a^{\dagger}_\sigma a_\sigma + \frac{1}{2} U_\mathrm{C}(x)n_\uparrow n_\downarrow
+ \frac{p^2}{2m}  + V_{N}(x).
\label{ham1}
\end{equation}
Here
$a^{\dagger}_\sigma$($a_\sigma $)  creates (annihilates) an electron with  spin~$\sigma=\uparrow,\downarrow$ in the molecular orbital and $n_\sigma = a^\dag_\sigma a_\sigma$.
The details of the parametrization of the molecular Hamiltonian are given in appendix A.
The model Hamiltonian is not restricted to the H$_2$ molecule.  It should rather be considered as a physically simple yet qualitative accurate  model of a covalent chemical
bond.

The complete Hamiltonian of the molecular junction consists of the molecular Hamiltonian (\ref{ham1}), the Hamiltonians for noninteracting left  and right electrodes, and the molecule-electrode
interaction:
\begin{equation}
H=H_M +  \sum_{\sigma, k \in L,R}  \varepsilon_k a^{\dagger}_{\sigma k} a_{\sigma k} +
\sum_{\sigma, k \in L,R } (t_k a^{\dagger}_{\sigma k} a_\sigma +\mbox{h.c.} ),
\label{ham2}
\end{equation}
where $a^{\dagger}_{\sigma k}$($a_{\sigma k}$) creates (annihilates) an electron in the single-particle state $\sigma k$ of either the left ($L$) or  the right ($R$) electrode.
The electron creation and annihilation operators satisfy standard fermionic anticommutation relations. The electron occupation numbers in both electrodes obey the Fermi-Dirac
distribution $f_{L,R}(\varepsilon_k) = [1 + e^{(\varepsilon_k-\mu_{L,R})/T}]^{-1}$.  The chemical potentials in the electrodes, $\mu_L$ and $\mu_R$, are assumed to be biased by the
external symmetrically applied voltage $V=\mu_L-\mu_R$, which we take to be positive, and $\mu_{L,R}=\pm0.5V$.
The Fermi energies of the electrodes are set to zero.
We also assume that the tunnelling  matrix element   $t_k$  is spin independent.

The coupling with the electrodes broadens the molecular level, with the width  given by the imaginary part of
the electrode self-energy,
\begin{equation}
\Gamma(\omega) =\Gamma_L + \Gamma_R=-\mathrm{Im}\sum\limits_{ k \in L,R }|t_{k}|^2/(\omega - \varepsilon_{k}+i0^+).
\end{equation}
In what follows, for the sake of simplicity,  we will use the wide-band  approximation for the  electrodes, i.e.,
the imaginary part $\Gamma$  of the self-energy is energy independent  and the real part vanishes. The total width of the molecular level is fixed in our
numerical calculations,  $\Gamma=1.36$ eV, but we vary the relative contributions of left and right electrodes.

We neglect the effect of the electric field on the molecule -- for  diatomic molecules lying flat on the metal surface  the electric field is  perpendicular to the bond length and does not influence the dynamics. In other cases, the electric field may give an additive contribution which can be easily included into the model for particular molecule-electrode geometries and inter-electrode distances.

We begin with the Ehrenfest coupled electron-nuclei dynamical equations
\begin{equation}
i \hbar \dot{\rho}(x,t) =[H,\rho(x,t)]
\end{equation}
\begin{equation}\label{Langevin0}
m \ddot x = -\mathrm{Tr}\bigl[\rho(x,t)  \frac{\partial H}{\partial x}  \bigr].
\end{equation}
We eliminate the electronic density matrix from this equation assuming that
the electronic degrees of freedom vary much faster than the nuclear motion. Physically this means that the oscillation period $\tau_p$ of the reaction coordinate
is much larger than the tunneling time for the electrons, i.e., $\tau_p\gg 1/\Gamma$. Below we will see that this assumption is fulfilled  for the model under consideration. This eliminates the direct time dependence from the electronic density matrix $\rho(x,t) \sim \rho(x)$, which now depends parametrically on time through the reaction coordinate.
The result is
the following Langevin equation for the reaction coordinate,
\begin{equation}\label{Langevin1}
m \ddot x = - \mathrm{Tr}\bigl[\rho(x) \frac{\partial H}{\partial x}\bigr]  -  \xi(x) \dot x + \delta f(x,t).
\end{equation}
Here $\rho(x)$ is the nonequilibrium density matrix, $\xi(x)$ is the frictional force (viscosity), and  $\delta f(x,t)$ is the random force (noise)
taken in the Gaussian form
\begin{equation}
\langle \delta f(x,t) \rangle=0,~~~~\langle \delta f(x,t) \delta f(x,t') \rangle =D(x) \delta(t-t'),
\label{fldis}
\end{equation}
The derivation of the Langevin equation via Keldysh nonequilibrium Green's functions  (NEGF) is presented in Appendix B.

\begin{figure}[t!]
 \begin{centering}
\includegraphics[width=1.0\columnwidth]{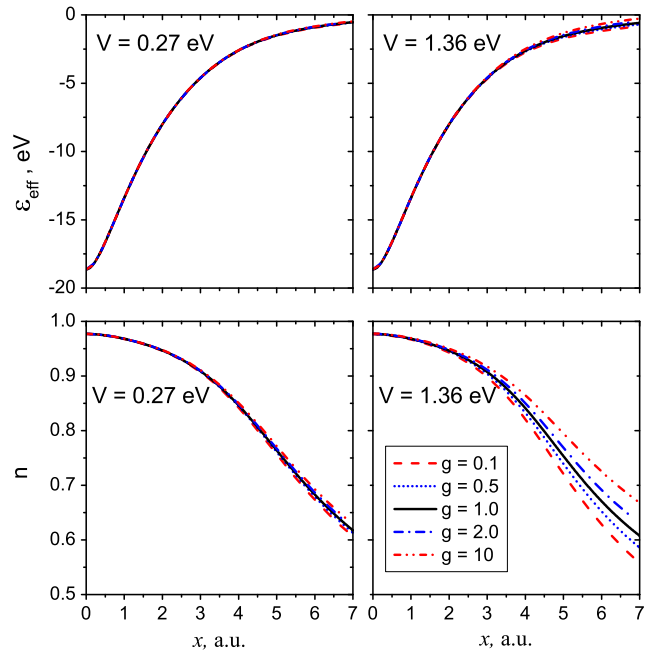}
\caption{Effective single particle energy level $\varepsilon_{\mathrm{eff}}(x)$ and population $n(x)$  as a function
of the reaction coordinate $x$ for different values of the asymmetry coefficient $g= \Gamma_L/\Gamma_R$ and applied voltage $V$. We chose $\varepsilon_\mathrm{eff}(x_\mathrm{min})=-9.8$ eV, where $x_\mathrm{min}=1.66$ a.u. corresponds to  the minimum of the equilibrium potential energy of the molecule. The Fermi energies of the electrodes are set to zero.}
\label{figure2} \end{centering}
\end{figure}

To obtain expressions for the frictional force and noise intensity $D(x)$ we apply the mean-field approximation to the Hamiltonian~(\ref{ham1}), i.e.,
we replace it by
\begin{equation}
H^\mathrm{mf}_M = \varepsilon_\mathrm{eff}(x)\sum_\sigma n_\sigma + V_N(x)+\frac{p^2}{2m},
\end{equation}
where
\begin{equation}\label{eps_eff}
\varepsilon_\mathrm{eff}(x)=\varepsilon(x) +V_g +\frac{1}{2} U_\mathrm{C}(x) n(x)
\end{equation}
is the effective single particle energy   and $n(x)=\mathrm{Tr}[\rho(x)n_\sigma]$ is the nonequilibrium population of the electronic level of the molecule.
This can be computed by means of the NEGF,  see Eq.~(\ref{ndef}) in Appendix B. In Appendix B we also present the details of the derivation of the explicit expressions for $\xi(x)$ and $D(x)$.

For low temperatures, $T\to0$, the Fermi-Dirac electron distributions in the electrodes $f_{L,R}(\omega)$ can be approximated by a step like function, and
the level population, friction coefficient, and noise intensity can be written as
 \begin{align}\label{n_x2}
n(x) = \frac12 -\frac{1}{\pi\Gamma}\left\{
 \Gamma_L\arctan\left(\frac{\varepsilon_\mathrm{eff}(x)-\mu_L}{\Gamma}\right) \right.
 \notag\\
    +
 \left. \Gamma_R\arctan\left(\frac{\varepsilon_\mathrm{eff}(x)-\mu_R}{\Gamma}\right)
 \right\}
\end{align}
\begin{equation}\label{ksi_x2}
  \xi(x) = \frac{2\Gamma}{\pi}\left\{\Gamma_L Q(\mu_L,x) + \Gamma_R Q(\mu_R,x) \right\},
\end{equation}
\begin{equation}\label{D_x2}
  D(x) = \frac{4\Gamma_L\Gamma_R}{\pi}\int_{\mu_R}^{\mu_L} d\omega \, Q(\omega,x),
\end{equation}
where the function $Q(\omega,x)$  is defined as
\begin{equation}\label{Qdef}
  Q(\omega,x) =  \left[\frac{\varepsilon'_\mathrm{eff}(x) }{(\omega-\varepsilon_\mathrm{eff}(x))^2 + \Gamma^2}\right]^2.
\end{equation}
Note that in the derivation of Eq.~\eqref{ksi_x2} we have utilized
 $(-\partial_\omega f_\alpha) =  f_\alpha(1-f_\alpha)/T$ and $\lim\limits_{T\to0}(-\partial_\omega f_\alpha)=\delta(\omega- \mu_\alpha)$.

Equations~(\ref{eps_eff}, \ref{ksi_x2}, \ref{D_x2}, \ref{n_x2}) are the main equations of our model. Their solution provides us with the
parameters of the Langevin equation~(\ref{Langevin1}). Note that $\varepsilon_\mathrm{eff}(x)$ in~(\ref{n_x2}) itself depends on $n(x)$, through Eq.~(\ref{eps_eff}). Therefore, both equations should be solved self-consistently. The main variables in  our model are the applied voltage $V$ and the asymmetry coefficient $g= \Gamma_L/\Gamma_R$.
The asymmetry coefficient $g$ controls the relative strength of the coupling to left and right electrodes. Replacing $g$ by $1/g$ is equivalent to reversing the applied voltage.

In Fig.~\ref{figure2} the effective single particle energy $\varepsilon_\mathrm{eff}(x)$ and the population $n(x)$ are depicted for two values of the applied voltage, small ($V=0.27$ eV) and large
($V=1.36$ eV),  and for different values of the asymmetry coefficient $g$.
The reference point for the effective molecular orbital energy is chosen to be $\varepsilon_\mathrm{eff}(x_\mathrm{min})=-9.8$ eV (relative to the electrode Fermi energy), where $x_\mathrm{min}=1.66$ a.u. determines the minimum of the $H_2$ ground state energy [see Eq.\eqref{H2_GrSt}].
This reference point can be shifted by the real part of the electrode self-energy, if we revoke the wide-band approximation, or  by the application of a gate voltage $V_g$. The dependence of the results on  gate voltage will be discussed below.
As  seen from the figure, at small values of the reaction coordinate, i.e., when the electron level is well below the electrode Fermi levels, both $\varepsilon_\mathrm{eff}(x)$
and $n(x)$ are nearly
independent of the applied voltage and the asymmetry  coefficient. The situation changes with increasing $x$, i.e., when $\varepsilon_\mathrm{eff}(x)$ comes into resonance
with the electrode Fermi levels. In that case, the electron population  is strongly affected by the asymmetry coefficient and the effect becomes more pronounced at larger values of $V$.
As a result, at large $x$ and $V$ the effective single particle energy  $\varepsilon_\mathrm{eff}(x)$  grows  with increasing $g$.

\begin{figure}[t!]
 \begin{centering}
\includegraphics[width=1.0\columnwidth]{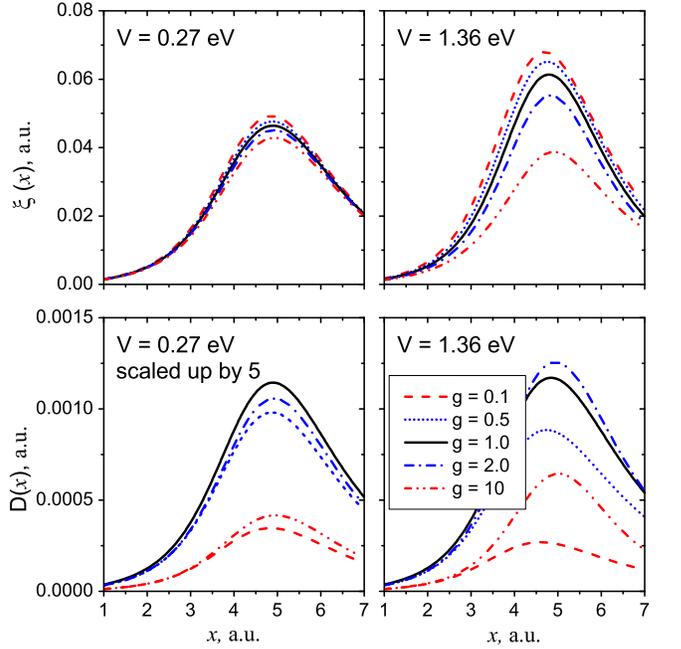}
\caption{ Friction coefficient $\xi(x)$ and noise intensity $D(x)$  as a function of the reaction coordinate $x$. All the parameters are the same as those in Fig.~\ref{figure2}. }
\label{figure3}
\end{centering}
\end{figure}

In Fig.~\ref{figure3} the friction coefficient $\xi(x)$ and the noise intensity $D(x)$ are plotted for the same parameters as those in Fig.~\ref{figure2}.
The friction coefficient grows  with increasing $x$ and reaches its maximum value when $\varepsilon_\mathrm{eff}(x)$ is close to the chemical potentials of  the electrodes,
i.e., when the resonant regime is approached. The friction $\xi(x)$ becomes smaller when the molecule starts to fall apart (due to the decrease of $\varepsilon'_\mathrm{eff}(x)$).
As seen from the figure, the friction coefficient grows as the asymmetry coefficient is reduced and this dependence
is more  pronounced for larger values of $V$. Besides, for small values of $g$ the friction is more sensitive to changes in the applied voltage.

The noise intensity $D(x)$ also reaches its maximum value when $\varepsilon_\mathrm{eff}(x)$ is close to the chemical potentials of the electrodes.
However, in contrast to the previous case, $D(x)$  is maximal for  symmetric coupling to the electrodes, i.e., when $g=1$.
Moreover,  because of the integral in Eq.~(\ref{D_x2}) the order of magnitude of $D(x)$ is proportional to the applied voltage.

\begin{figure}[t!]
 \begin{centering}
\includegraphics[width=1.0\columnwidth]{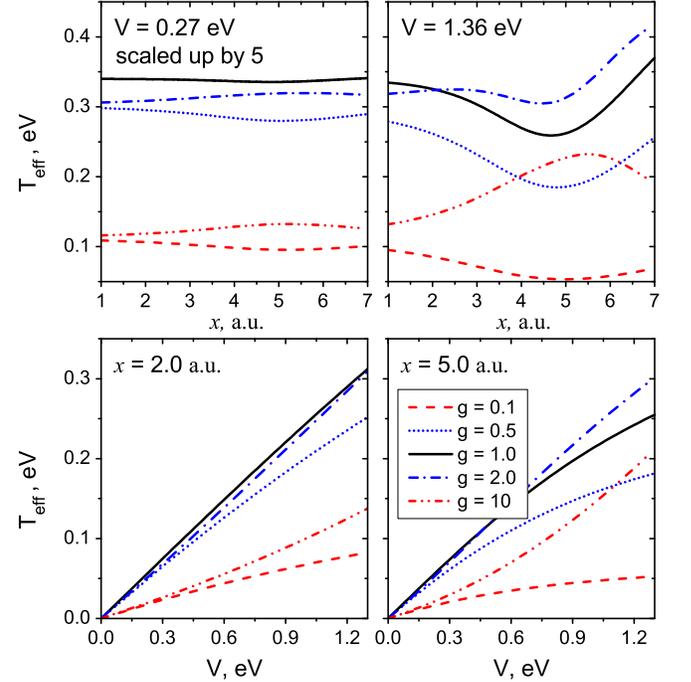}
\caption{Upper panels: Effective temperature $T_\mathrm{eff}$ as a function of the reaction coordinate $x$ for two values of the applied voltage $x$.
Lower panels: Effective temperature $T_\mathrm{eff}$ as a function of the applied voltage $V$ for two values of the reaction coordinate $x$.
All the parameters are the same as those in Fig.~\ref{figure2}.}
\label{Tloc_x} \end{centering}
\end{figure}

Motivated by the fluctuation-dissipation theorem, we can also introduce an effective "temperature" which depends on the reaction coordinate,
\begin{equation}\label{T_eff1}
  T_\mathrm{eff}(x) = \frac{D(x)}{2\xi(x)} =
  \frac{\Gamma^{-1} \Gamma_R \Gamma_L \int_{\mu_R}^{\mu_L}Q(\omega,x)d\omega}{\Gamma_L Q(\mu_L,x) + \Gamma_R Q(\mu_R,x)}.
\end{equation}
Assuming that $Q(\omega,x)\approx\mathrm{const}$ for $\mu_L > \omega > \mu_R$ we can approximate~(\ref{T_eff1}) by
\begin{equation}\label{T_eff2}
  T_\mathrm{eff}(x) \approx T(V,g)=V \frac{\Gamma_R \Gamma_L}{\Gamma^2}= V \frac{g}{(1+g)^2},
\end{equation}
The effective temperature~\eqref{T_eff2}  has a maximum of  $V/4$ when $g=1$. Note that the current is also maximal when the coupling to the  electrodes is symmetric.
In the upper panels of Fig.~\ref{Tloc_x} the effective temperature $T_\mathrm{eff}(x)$ is shown for the same  parameters as those in Fig.~\ref{figure2} for two values of the applied voltage.
As seen, for small applied voltages $T_\mathrm{eff}(x)$ is nearly independent of $x$ and  can be approximated by Eq.~\eqref{T_eff2}.
When  the applied voltage increases the effective temperature also grows and $T_\mathrm{eff}$ becomes dependent on
the reaction coordinate  in a rather nontrivial way. The lower panels of Fig.~\ref{Tloc_x} depict the effective  temperature as a function of $V$ for two particular values
of the reaction coordinate. The point $x=2.0$ a.u. is close to the minimum of the effective potential, while $x=5.0$ a.u. is near the top of the barrier. We emphasize
that by reversing the applied voltage polarity (i.e.,  replacing $g$ by $1/g$) we can vary the effective temperature and the effect is more pronounced  the larger the value of $V$.

\section{Reaction rates}
\subsection{ Fokker-Planck equation for the reaction coordinate}

\begin{figure}[t!]
\begin{centering}
\includegraphics[width=1.0\columnwidth]{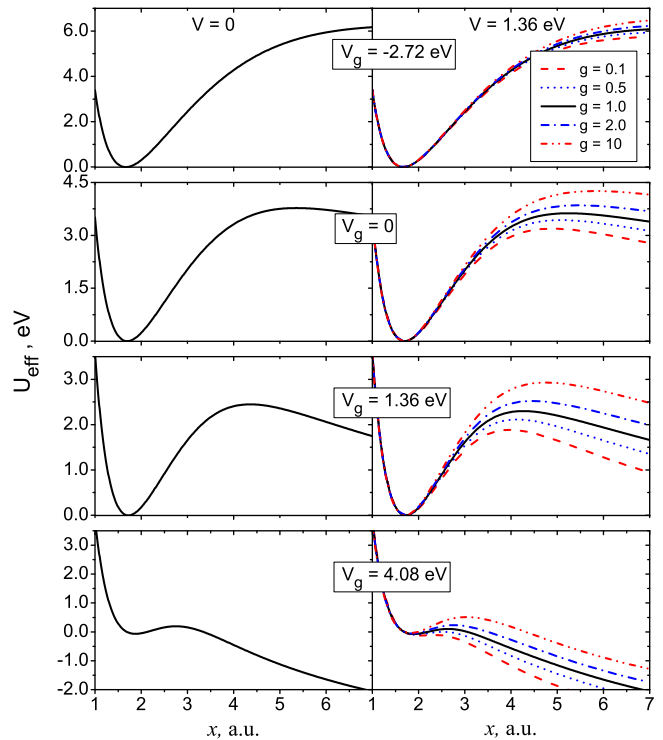}
\caption{Effective nonequilibrium potential $U_{\mathrm{eff}}(x)$ as a function of the reaction coordinate $x$
for two values of the applied voltage $V$. The panels correspond to different positions of the effective molecular orbital energy relative to the electrode Fermi energy $\varepsilon_\mathrm{eff}(x_\mathrm{min})=-9.8 \mbox{ eV} +V_g$. All other parameters are the same as those in Fig.~\ref{figure2}.}
\label{figure5}
\end{centering}
\end{figure}

Since  $\partial H/\partial x = \varepsilon'_\mathrm{eff}(x)\sum_\sigma n_\sigma + V'_N(x)$ in  mean-field approximation, Eq.~\eqref{Langevin1} becomes
\begin{equation}\label{Langevin2}
m \ddot x = - 2\varepsilon'_\mathrm{eff}(x)n(x)- V'_N(x)  -  \xi(x) \dot x + \delta f(x,t).
\end{equation}
The Langevin equation~\eqref{Langevin2} is equivalent to the Fokker-Planck equation for the distribution function $F(x,p,t)$,
\begin{align}\label{FP}
\frac{\partial}{\partial t}  F = - \left\{U'_\mathrm{eff}(x)\frac{\partial}{\partial p} - \frac{p}{m} \frac{\partial}{\partial x}\right\}F
\notag\\
+ \frac{\xi(x)}{m} \frac{\partial}{\partial p}(p F) + \frac{D(x)}{2} \frac{\partial^2}{\partial p^2} F.
 \end{align}
Here the effective nonequilibrium potential energy surface is defined via  integration of the
 nonequilibrium force in  the Langevin equation,
 \begin{equation}
 U_\mathrm{eff}(x) = V_N(x) +   2\int_{x_0}^{x} \varepsilon'_\mathrm{eff}(y) n(y) dy.
 \end{equation}

In Fig.~\ref{figure5} the nonequilibrium effective potential is shown as a function of the reaction coordinate
for different values of the asymmetry coefficient $g$, applied voltage bias $V$, and gate voltage $V_g$. The current flow through the molecule changes
the height of the potential barrier: for $g<1$ the current reduces the barrier height and for $g>1$ the current increases it.
The explanation is the following: when the coupling to  the left electrode is stronger ($g>1$) the current pumps electrons into the molecule
and enhances the chemical bond. Conversely, when the coupling to the right electrode is stronger the  current depletes the molecular electrons, thereby
weakening the chemical bond. Thus, by reversing the voltage polarity we can vary the height of the potential barrier.  But it should be stressed
that the effective temperature $T_\mathrm{eff}$ also changes with reversing the voltage polarity and the barrier growth is accompanied by a temperature increase.
We also see from Fig.~\ref{figure5} that the height of the potential barrier can be decreased by increasing the gate voltage.
The resonant tunneling regime $V_g =0$ and $V_g=1.36$ eV is physically most  interesting, since the nonequilibrium potential energy profile exhibits a clear barrier between product and reactant states. We focus on the resonant tunnelling regime in our calculations of the reaction rates.

\begin{table}[t!]
\begin{centering}
\begin{tabular}{cccccccc}
\hhline{========}
   &\multicolumn{3}{c}{ $V_g=0.0$ }\rule[-4pt]{0cm}{14pt}  & & \multicolumn{3}{c}{ $V_g=1.36$ eV }
\\
\hhline{~---~---}
    ~$k$~ & $g=0.1$ & $g=1.0$ & $g=10$ & ~~ & $g=0.1$ & $g=1.0$ & $g=10$\rule[-4pt]{0cm}{14pt} \\
\hhline{--------}
  $0.1$ & 400     &   405   & 393  & & 435     & 414     & 407\rule[-0pt]{0cm}{12pt}       \\
  $0.5$ & 498     &   496   & 497  & & 515     & 500     & 491\rule[-0pt]{0cm}{12pt}       \\
  $0.9$ & 781     & 806     & 817  & & 744     & 752     & 776\rule[-0pt]{0cm}{12pt}  \\
\hhline{========}
\end{tabular}
\caption{The period of motion (Eq.~\eqref{T_E},  a.u. of time) in the nonequilibrium
effective potential (the bias voltage $V = 1.36$ eV) for different values of the
asymmetry coefficient $g$ and the gate voltage $V_g$. The parameter $k$
defines the energy, $k = (E - E_\mathrm{min})/ (E_\mathrm{max} -
E_\mathrm{min})$.  All other parameters are the same as those in Fig.~\ref{figure2}. }
\label{periodd}\end{centering}
\end{table}

Now we want to reduce the Fokker-Planck equation~\eqref{FP} to the Smoluchowski equation for the distribution function which depends either on the reaction coordinate $x$ (overdamped motion)
or on the energy $E = \frac{p^2}{2m} + U_\mathrm{eff}(x)$ (underdamped motion). To find which case is realized for the considered model we compare the oscillation period
along the reaction coordinate and  the relaxation time due to friction. The oscillation period is given by
\begin{equation}
  \tau_p(E) = 2\int_{x_1}^{x_2} \frac{dx}{\dot{x}} = 2\int_{x_1}^{x_2} \frac{dx}{\sqrt{\frac{2}{m}(E- U_\mathrm{eff}(x))}},
  \label{T_E}
\end{equation}
where $x_{1}$ and $x_2$ are left and right turning points, i.e., $E= U_\mathrm{eff}(x_{1,2})$. If $E\approx U_\mathrm{eff}(x_\mathrm{min})$ then
$\tau_p\approx 2\pi/\omega_0$, where $\omega_0 = \sqrt{ U''_\mathrm{eff}(x_\mathrm{min}) /m}$ is the oscillation frequency near the bottom of the effective potential.
In Table~\ref{periodd} the
oscillation period is computed for various values of the bias voltage, gate voltage, and  asymmetry coefficient. Since $\tau_p\gg 1/\Gamma$ ($\Gamma = 0.05$ a.u.)
the assumption behind the derivation of the Langevin equation (appendix B), namely the assumption that the  electronic degrees of freedom are much faster than the motion along
$x$ is fulfilled. In Fig.~\ref{figure3} the friction $\xi(x)$ is shown for different values of asymmetry coefficient $g$ and applied voltage $V$.
It is evident that the relaxation time due to the friction $m/\xi(x)$  is much larger than the oscillation period. Therefore the energy dissipation per period of the motion is small, which means that we deal with {\it underdamped  motion}.

\subsection{Calculations of the reaction rates }

The solution of  the Fokker-Planck equation  is not trivial in our case, since the effective temperature depends on the reaction coordinate.
Using the method described by~Coffey {\em et al.} {\cite{coffey} the Fokker-Planck equation is reduced to the equation for the
distribution function for the energy, $P(E,t)$
\begin{align}
\frac{\partial}{\partial t} P(E,t) = & \frac{\partial}{\partial E} \left[\xi(E) P(E,t)\right]
\notag\\
 &+\frac12 \frac{\partial}{\partial E} \left[D(E) \frac{\partial}{\partial E}P(E,T)\right],
\end{align}
where
\begin{equation}
  \xi(E) = \frac{1}{T(E)}\int_{x_1}^{x_2} \xi(x)\dot{x} dx
\end{equation}
and
\begin{equation}
  D(E) = \frac{1}{T(E)}\int_{x_1}^{x_2} D(x)\dot{x} dx.
\end{equation}
By introducing
\begin{equation}\label{alpha}
 \alpha(E) = \frac12  \frac{\partial}{\partial E}D(E) - \xi(E)
\end{equation}
we re-write  this equation as the Smoluchowski equation
\begin{equation}
\frac{\partial}{\partial t} P(E,t) =  - \frac{\partial}{\partial E} \left[\alpha(E) P(E,t)\right] + \frac12 \frac{\partial^2}{\partial E^2}  \left[D(E) P(E,T)\right].
\end{equation}
The same equation was obtained in Ref.~\onlinecite{PhysRevB.73.035104}, though the theory  was applied only to a nano-mechanical harmonic oscillator and thus did not consider dissociation processes.
Then, the mean time for a particle with initial energy $E_\mathrm{min}=U_\mathrm{eff}(x_\mathrm{min})$
to arrive at the top  $E_\mathrm{max}=U_\mathrm{eff}(x_\mathrm{max})$ of the potential barrier is given by (see, for example, Eq.~XII.3.7 in Ref.~\onlinecite{vanKampen})
\begin{equation}\label{tau1}
  \tau = 2 \int_{E_\mathrm{min}}^{E_\mathrm{max}} dE e^{\Phi(E)} \int^E_{E_\mathrm{min}}  \frac{dE'}{D(E')}e^{-\Phi(E')},
\end{equation}
where
\begin{equation}
 \Phi(E) = - \int_{E_0}^E \frac{2\alpha(E')}{D(E')} dE'.
\end{equation}
Note, that the choice of $E_0$ is not relevant since it does not contribute to the escape time~\eqref{tau1}.
Using relation~\eqref{alpha} we can rewrite~\eqref{tau1} as
\begin{equation}\label{tau2}
 \tau = 2 \int_{E_\mathrm{min}}^{E_\mathrm{max}}\frac{dE}{D(E)}   e^{\Psi(E)} \int^E_{E_\mathrm{min}} e^{-\Psi(E')}dE' ,
\end{equation}
where
\begin{equation}\label{psi_E}
 \Psi(E) =   \int_{E_0}^E \frac{1}{T_\mathrm{eff}(E')} dE'
\end{equation}
in terms of the effective temperature $T_\mathrm{eff}(E) = {D(E)}/2\xi(E)$.

Equation~\eqref{tau2} gives us the exact escape time (mean first passage time) for the Brownian particle in the underdamped regime with coordinate-dependent effective temperature. In our particular case it can be further simplified,
because our calculations show that the effective temperature $T_\mathrm{eff}(E)$ depends only slightly
on energy when $E_\mathrm{min}< E < E_\mathrm{max}$ and can be approximated by Eq.~\eqref{T_eff2}., i.e., $T_\mathrm{eff}(E)=T(V,g)$. With this approximation,
the integral~\eqref{psi_E} becomes $\Psi(E) \approx  E/T_\mathrm{eff}(E)$. Substituting this  into Eq.~\eqref{tau2} we obtain
\begin{equation}\label{tau3}
  \tau\approx 2 T_\mathrm{eff}(E_\mathrm{min}) \int_{E_\mathrm{min}}^{E_\mathrm{max}}\frac{dE}{D(E)}
  \exp\left\{\frac{E-E_\mathrm{min}}{T_\mathrm{eff}(E)} \right\},
\end{equation}
This formula gives us the escape time from the nonequilibrium potential barrier with the effective temperature which depends on the energy of the particle.

\begin{figure}[t!]
\begin{center}
\includegraphics[width=1.0\columnwidth]{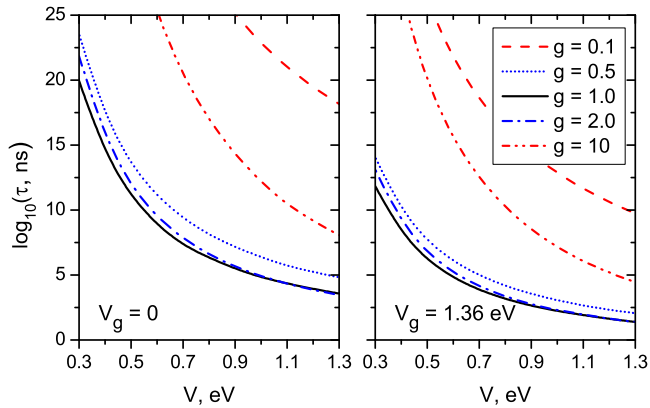}
\end{center}
\caption{Escape time computed for various values of the asymmetry coefficient $g$ as a function of the applied voltage. The left and right panels correspond to different positions of the effective molecular orbital energy relative to the electrode Fermi energy $\varepsilon_\mathrm{eff}(x_\mathrm{min})=-9.8 \mbox{ eV} +V_g$. All other parameters are the same as those in Fig.~\ref{figure2}.}
\label{esc_time}
\end{figure}

In Fig.~\ref{esc_time} we show the escape time~\eqref{tau3} computed for various values of the asymmetry coefficient $g$ as a function of the applied voltage $V$.
For all values of $g$ the escape time decreases with increasing applied voltage. We also see that moving the molecular orbital closer to the Fermi energy of the electrodes facilitates the dissociation.
When the noise is balanced by the viscosity, i.e. when the fluctuation-dissipation relations is forced for the nuclear degrees of freedom,\cite{dzhioev11}
the dependence of the reaction rates on asymmetry coefficient is more complicated. For example, it was shown that the chemical reaction can be catalysed or stopped depending on the direction of the electric current.\cite{dzhioev11} In contrast,  the absence of the fluctuation-dissipation relation leads to a rise of the effective temperature $T_\mathrm{eff}(x)$ (or $T_\mathrm{eff}(E)$) which always overrides the increase of the nonequilibrium potential barrier. We also see from the figure that for asymmetric coupling the escape time  can be controlled
 by changing the applied voltage polarity, i.e., by replacing $g$ by $1/g$.

\begin{figure}[t!]
\begin{center}
\includegraphics[width=1.0\columnwidth]{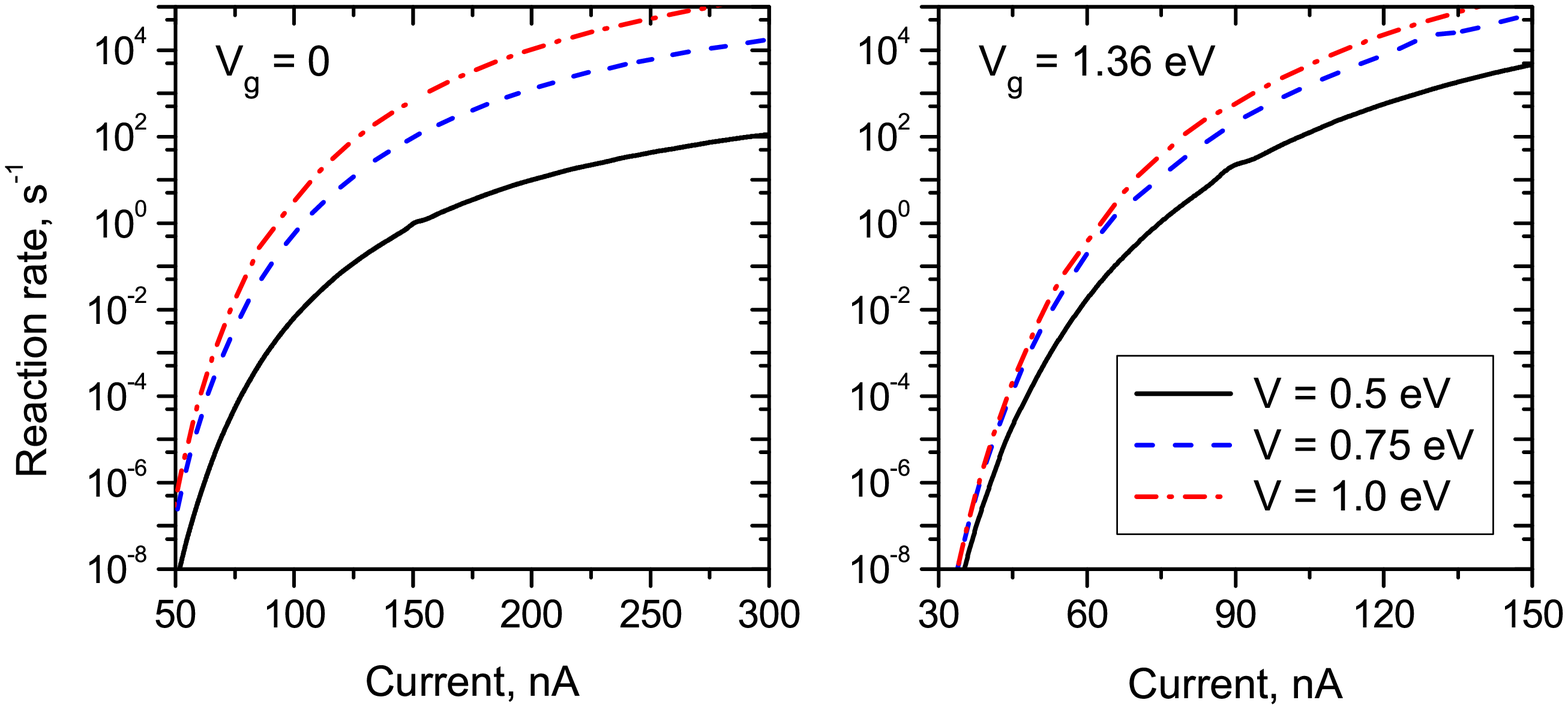}
\end{center}
\caption{Reaction rate for the molecule dissociation computed at the fixed applied voltage as a function of the electric current. The left and right panels correspond to different positions of the effective molecular orbital energy relative to the electrode Fermi energy $\varepsilon_\mathrm{eff}(x_\mathrm{min})=-9.8 \mbox{ eV} +V_g$. In our calculations we choose the right coupling  fixed, $\Gamma_R=1.36$~eV, and vary the left coupling to change the current at  constant applied voltage. All other parameters are the same as those in Fig.~\ref{figure2}.}
\label{reaction_rate}
\end{figure}

In experiments on STM induced molecular dissociation, the applied voltage bias is usually fixed while the electric current is varied by  changing the distance between STM tip and molecule.\cite{PhysRevLett.78.4410}
In the following, we apply our approach to  model this  experimental scenario. Similar to the experimental conditions, we keep  the coupling to the surface, $\Gamma_R$, and the applied bias voltage fixed. Then we compute the reaction rate (inverse  escape time) as a function of the electric current by changing the coupling to the left electrode $\Gamma_L$, which is equivalent to changing the distance between  STM tip and molecule. The results of the calculation are shown in Fig.\ref{reaction_rate}. Our calculations predict a rather nontrivial dependence of the reaction rate on the tunnelling current.
Note that these results reveal much more structure than the power
law dependence on current reported experimentally for a related but different
system, namely O$_2$.\cite{PhysRevLett.78.4410}  The main reason for this
is the following.
The simple estimation shows that the coupling to the STM tip, $\Gamma_L$, which reflects the experimental conditions,\cite{PhysRevLett.78.4410}  is  of order $ 0.001 $ eV. This makes the
electronic tunneling rate small compared to the vibrational frequency
of the molecule. In this limit, a simple Fermi Golden Rule calculation
applies and gives a power law dependence of the reaction rate.\cite{PhysRevLett.78.4410}
In contrast, we treat the interesting limit of larger electronic currents where the molecule is
in fully developed nonequilibrium. It is in this regime that we predict a
highly nontrivial current dependence of the dissociation rate.
This reflects the fact that our approach is geared towards fully developed
nonequilibrium, while previous work effectively treated systems close to
thermal equilibrium.
Therefore, the absence of a simple power low dependence of the  reaction rate  and the appearance of  nontrivial nonlinear behaviour as shown in Fig.\ref{reaction_rate}
may serve as an indication of  fully developed electronic nonequilibrium.

\section{Conclusions}

In this paper, we have formulated and solved Kramers problem for current-induced, out of equilibrium chemical reactions.  As a  general model of covalent bond breaking, we  have considered the dissociation of a diatomic molecules induced by the tunnelling  current from a STM tip. We have  proposed a model Hamiltonian for the system and, by projecting out the electronic degrees of freedom, we  have derived a Langevin equation for the time evolution of the reaction coordinate. The Langevin equation leads to  Fokker-Planck dynamics with an effective temperature which depends on the reaction coordinate. The reaction rate  for the dissociation is computed by solving the Fokker-Planck equation for the escape time.
In the resonant tunnelling regime, two processes play equally important roles in the molecular dissociation: the decrease of the potential energy barrier and reaction-coordinate-dependent local heating of the molecule. The current induced catalytic reduction of the barrier accompanied by the local heating is an alternative mechanism for current-induced bond breaking, which is very different from the widely accepted paradigm of  merely pumping vibrational degrees of freedom.

\begin{acknowledgments}
The authors thank M. Gelin for many valuable discussions.
 This work has been supported by the Francqui Foundation, and
 Programme d'Actions de Recherche Concert\'ee de la Communaut\'e Francaise, Belgium as well as
 SFB 658 of the Deutsche Forschungsgemeinschaft.

\end{acknowledgments}

\appendix
\section{Parametrization of the Hamiltonian}
In our calculations we take $\varepsilon(x)$ in the form of the bonding
orbital for the $H_2^+$ molecule in the  gas phase~\cite{mcquarrie-qc}
\begin{equation}
 \varepsilon(x) = -\frac{1}{2} + \frac{J(x) + K(x)}{1 + S(x)}
\end{equation}
where the functions $J(x),~K(x),~S(x)$ are given by
\begin{align}
 J(x) &= \mathrm{e}^{-2x}\left(1+\frac1x\right) - \frac1x,~~K(x)= - \mathrm{e}^{-x}(1+x),
\notag\\
S(x) &= \mathrm{e}^{-x}\left(1 + x + \frac{x^2}{3}\right).
\end{align}
The Coulomb interaction between electrons  $U_\mathrm{C}(x)$ is chosen such to reproduce, within the equilibrium ($n(x)=1,~V_g=0$) mean-field Hamiltonian~\eqref{ham2}, the
energy of the $H_2$ molecule obtained within the molecular orbital theory (see Appendix on page 543 in Ref.~\onlinecite{mcquarrie-qc}), i.e.,
\begin{equation}\label{H2_GrSt}
 \braket{H^\mathrm{mf}_M} = E_{MO}(x) =2\varepsilon(x) +  U_\mathrm{C}(x)+\frac{1}{x}.
\end{equation}
Here we neglect the nuclear kinetic energy contribution.
Therefore, $U_\mathrm{C}(x)$ can be written as follows
\begin{equation}
  U_\mathrm{C}(x) = \frac{\frac{5}{16}+0.5J_p(x)+ K_p(x)+2L(x)}{(1+S(x))^2},
\end{equation}
where the functions $L(x),~J_p(x),~K_p(x)$ are listed in Table~10.8 of Ref.~\onlinecite{mcquarrie-qc}.

\section{Derivation of the Langevin equation for the reaction coordinate via NEGF}

The Langevin equation for nuclear degrees of freedom coupled to out-of-equilibrium electrons has recently been given in various
 contexts~\cite{PhysRevB.73.035104,Pistolesi08,Bode11,Bode12,dundas12,PhysRevB.86.195419}. To make this paper self-contained, we provide here an explicit derivation which is directly adapted to the Hamiltonian under consideration. Our starting point is the Heisenberg equation of motion for the reaction coordinate as obtained from the mean-field Hamiltonian (\ref{ham2}),
 \begin{equation}
   m \ddot{x} + \frac{\partial V_N}{\partial x} = - \varepsilon'_{\rm eff}(x)\sum_{\sigma} a^\dagger_\sigma a_{\sigma}\, ,
   \label{HEOM}
\end{equation}
Here, the right-hand side contains the operator describing the current-induced forces. Given that the molecular dissociation dynamics is slow compared to the electronic degrees of freedom, we calculate these forces within a nonequilibrium adiabatic approximation. In this approximation, the reaction coordinate is taken as classical. The electronic state should be computed for a given trajectory $x=x(t)$ while in turn, the electrons affect this trajectory through adiabatic reaction forces.

In the  nonequilibrium adiabatic approximation, one averages the force appearing on the right-hand side of Eq.\ (\ref{HEOM}) over times which are long on the scale of the  electronic dynamics, but short on the scale of the dissociation dynamics. The average adiabatic reaction force is thus given by the expectation value $\langle \varepsilon^\prime_{\rm eff}(x) a_\sigma^\dagger a_\sigma\rangle_{x(t)}$, evaluated for a given trajectory ${x}(t)$ of the reaction coordinate. The fluctuation-dissipation theorem implies that we need to include a fluctuating Langevin force $\delta f(t)$ in addition to the friction force. Hence Eq. \eqref{HEOM} becomes
\beq
 m \ddot x + \frac{\partial V_N}{\partial x} = 2i \varepsilon^{\prime}_{\rm eff}(x)
{\cal G}^<(t,t)+\delta f(t) ,
   \label{NEBO}
\eeq
where we have introduced the lesser Green's function
\beq\label{GlessDef}
{\cal G}^<(t,t')= i\langle a_{\sigma}^\dagger(t')a_\sigma^{}(t)\rangle_{{ x}(t)}\,,
\eeq
and the factor of two accounts for spin. Thus, except for the stochastic noise force, the
current-induced forces are encoded in $2i \varepsilon^{\prime}_{\rm eff}(x)
{\cal G}^<$. Below we demonstrate that the current-induced forces can be represented as a sum of two contributions:
the adiabatic force and a velocity-dependent contribution,
\begin{equation}
  2i \varepsilon^{\prime}_{\rm eff}(x){\cal G}^< \simeq F(x) - \xi(x) \dot x.
\end{equation}

The variance of the stochastic force $\delta f(t)$ is governed by the symmetrized fluctuations of the operator $\varepsilon^\prime_{\rm eff}(x) a_\sigma^\dagger a_\sigma$. Given that the electronic fluctuations happen on short time scales,  $\delta f(t)$ is locally correlated in time,
\beq\label{defvariance}
\langle \delta f(t) \delta f(t')\rangle= D({x}) \delta(t-t')\,.
\eeq
Since we are dealing with a mean-field Hamiltonian, $D({x})$ can be evaluated using Wick's theorem,
\beq \label{xixi}
\langle \delta f(t) \delta f(t')\rangle  = 2 [\varepsilon^\prime_{\rm eff}(x)]^2 {\cal G}^>(t,t^{\prime}) {\cal G}^<(t^{\prime},t)\,,
\eeq
where
\begin{align}\label{eq:G>}
{\cal G}^{>}(t,t')=-i \langle a_{\sigma}(t) a_{\sigma}^{\dagger}(t') \rangle_{{x}(t)}\,
\end{align}
is the greater Green's function. These expressions show that we need to evaluate the electronic Green's function for a given classical trajectory ${x}(t)$.

We start with the Keldysh equation for lesser Green's function:
\begin{align}\label{eq:Gddless}
  {\cal G}^<(t,t^{\prime})=
\int\mathrm{d}t_1 \int\mathrm{d}t_2\, {\cal G}^R(t,t_1)
\Sigma^<(t_1,t_2)  {\cal G}^A(t_2,t^{\prime})\,,
\end{align}
where the retarded and advanced Green's functions are given by the standard expressions
\beq\label{GRdef}
{\cal G}^R(t,t^{\prime}) =
-i \theta(t-t') \langle \{a_\sigma (t), a_\sigma^{\dagger}(t^{\prime})
\}\rangle_{{x}(t)}\,,
\eeq
\beq\label{GAdef}
{\cal G}^A(t,t^{\prime}) = {\cal G}^R(t^{\prime},t)^*.
\eeq
The Keldysh equation (\ref{eq:Gddless}) involves the lesser self-energy
\beq\label{SelfEnL}
\Sigma^<(\omega) =2
i\, \sum_{\alpha} f_{\alpha}(\omega) \Gamma_{\alpha}(\omega)\,,
\eeq
where $f_{\alpha}(\omega)=[1 + e^{\beta_\alpha(\omega - \mu_\alpha)}]^{-1}$ is the Fermi-Dirac electron distribution
in the left ($\alpha = L$) or the right ($\alpha = R$) electrode, and
 $\Gamma_\alpha(\omega)$ is the imaginary part of the retarded self-energy
due to interaction with the electrodes
\begin{equation}
\Sigma^R_\alpha(\omega) = \sum\limits_{k\in\alpha}\frac{|t_{k}|^2}{\omega - \varepsilon_{k}+i0^+}=\Lambda_\alpha(\omega)-i\Gamma_\alpha(\omega).
\end{equation}

The adiabatic expansion of Keldysh equation is carried out in the Wigner representation, given by
\begin{equation}
\tilde{A}(t,\omega)=\int\mathrm{d}\tau\,e^{i\omega\tau} A(t+\tau/2,t-\tau/2)
\end{equation}
for a general function $A(t,t^\prime)$, in which fast and slow time scales are easily identifiable. For the Green's function ${\cal G}^{A,R,<}$, the slow mechanical motion implies that ${\cal G}^{A,R,<}(t_1,t_2)$ varies slowly with the central time $t=\frac{t_{1}+t_{2}}{2}$, but oscillates fast with the relative time $\tau=t_{1}-t_{2}$. As usual, the Wigner transform of a convolution $C(t_{1},t_{2})=\int\mathrm{d}t_{3}\, A(t_1,t_{3})B(t_{3},t_{2})$
is given by
\begin{eqnarray}\label{moyalexp}
\tilde{C} & = &
\exp\left[\frac{i}{2}\left(\partial_{\epsilon}^{\tilde{A}}\partial_{t}^{\tilde{B
}}
-\partial_ { t } ^{\tilde{A}}\partial_{\epsilon}^{\tilde{B}}\right)\right]\tilde
A\tilde B\nonumber \\
 & \simeq & \tilde A\tilde
B+\frac{i}{2}\partial_{\epsilon}\tilde
A\partial_{t}\tilde
B-\frac{i}{2}\partial_{t}\tilde
A\partial_{\epsilon}\tilde
B,
\end{eqnarray}
where we have dropped higher order derivatives in the last line, exploiting the slow variation with the central time $t$.

Expanding Eq. \eqref{eq:Gddless} up to the leading adiabatic correction according to Eq. \eqref{moyalexp} and taking into account that  $\Sigma^<$ depends only on $\omega$ and is independent of the central time, we obtain ${\cal G}^<$ to first order in $\dot{x}$,
\begin{align}\label{GLadexp}
{\cal \tilde{G}}^< =& G^< +\frac{i}{2} \, \dot x \,  \varepsilon^\prime_{\rm eff}(x) \,
\left[G^< \partial_\omega G^> +  G^> \partial_\omega G^<\right]
\end{align}
Here ${\cal \tilde G}$ denotes  full Green's
functions in Wigner representation, while $G$ denotes the strictly adiabatic (or "frozen") Green's functions  that are evaluated for a fixed value of $x$:
$G^{R}(x,\omega)=\left[\omega -\varepsilon_{{\rm eff}}(x) -\Sigma^{R}(\omega) \right]^{-1}$,
$\Sigma^R(\omega)= \sum_\alpha \Sigma^R_\alpha(\omega)$,
${G}^A=({ G}^R)^\dagger$, $G^<=G^R\Sigma^<G^A $,  $G^>= G^< + G^R - G^A $.

Let us now compute the current-induced forces appearing in the Langevin equation \eqref{Langevin1}.  In the strictly
adiabatic limit, {\it i.e.}, retaining only the first term on the RHS of Eq.~(\ref{GLadexp}), ${\cal \tilde G}^<
\simeq G^<$, we obtain the mean force
\beq\label{fdef}
F(x) = - 2\varepsilon^{\prime}_{\rm eff}(x)\, n(x)\,
\eeq
where $n(x)$ is the nonequilibrium population of the electronic level
\begin{equation}\label{ndef}
  n(x) = \int\frac{d\omega}{2\pi i}\,G^< (x,\omega).
\end{equation}

The leading order correction in
Eq.~(\ref{GLadexp}) gives a velocity-dependent contribution to the current induced forces, which determines the friction $\xi(x)$ in the Langevin equation. After integration by parts
we obtain the explicit expression
\begin{equation}\label{xidef}
\xi(x) = 2[\varepsilon^{\prime}_{\rm eff}(x)]^2  \int \frac{d\omega}{2\pi }  \, G^< (x,\omega)\partial_\omega G^> (x,\omega).
\end{equation}
We evaluate the noise intensity $D(x)$ (\ref{xixi}) for the stochastic force
$\delta f$  to the lowest order in the adiabatic expansion, so that
\beq \label{ddef}
D(x)= 2[\varepsilon^\prime_{\rm eff}(x)]^2 \int \frac{d\omega}{2\pi }\,  { G}^>(x,\omega) { G}^<(x,\omega).
\eeq

Expressions Eqs,~(\ref{xidef},\ref{ddef}) can be simplified if we use the wide-band approximation for the electrodes. Namely, the
wide-band limit employs that the retarded self-energy is energy independent, $\Sigma_\alpha^R(\omega) = -i\Gamma_\alpha$.
In that case
\begin{equation}
  G^< (x,\omega) = 2i \frac{\sum_\alpha f_\alpha \Gamma_\alpha}{ (\omega - \varepsilon_\mathrm{eff}(x))^2 + \Gamma^2}
\end{equation}
and we  we obtain the following expressions for the friction
\begin{equation}\label{xidef2}
  \xi(x) = 4\Gamma\int \frac{d \omega}{2 \pi}\,Q(\omega, x)  \sum_{\alpha = L, R} (-\partial_\omega f_\alpha) \Gamma_\alpha
\end{equation}
and the noise intensity
\begin{equation}\label{Ddef2}
D(x)  = 8\int \frac{d \omega}{2 \pi}\, \,Q(\omega, x)
\sum_{\alpha \alpha' = L,R} \Gamma_{\alpha} \Gamma_{\alpha'} f_{\alpha}(1- f_{\alpha'}).
\end{equation}
Here,  the factor $Q(\omega, x)$ is given by Eq.~\eqref{Qdef}.

\end{document}